\documentclass[letterpaper,10pt,conference]{ieeeconf}
\newcommand{\ONLINEVERSION}{}

\IEEEoverridecommandlockouts
\usepackage{cite}
\usepackage{amsmath,amssymb,amsfonts}
\usepackage{algorithmic}
\usepackage{graphicx}
\usepackage{textcomp}
\usepackage{enumerate}

\usepackage{xcolor}
\usepackage{subcaption}
\usepackage{bm}
\usepackage{arydshln}
\usepackage[implicit=false, colorlinks=true, urlcolor=blue]{hyperref}

\newcommand{\expectation}{\mathbf{E}}
\newcommand{\expectationp}[1]{\mathbf{E}\left[ {#1} \right]}

\newcommand{\expectationbelowp}[2]{\underset{#1}{\mathbf{E}}\left[ {#2} \right]}

\newcommand{\variancep}[2][]{\mathbf{V} \ifx #1 \undefined \else _{#1} \fi \left[#2\right]}

\newcommand{\covariancep}[2][]{\mathbf{K} \ifx #1 \undefined \else _{#1} \fi \left[#2\right]}

\newcommand{\kullbp}[2]{D_{\text{KL}} \left( {#1} \, \lVert \, {#2} \right)}

\newcommand{\condbar}{\,|\,}  

\newcommand{\normaldist}{\mathcal{N}}

\newtheorem{prop}{Proposition}
\newtheorem{thm}{Theorem}

\newtheorem{lem}{Lemma}

\newtheorem{cor}{Corollary}

\DeclareMathOperator*{\argmin}{argmin} 
\DeclareMathOperator*{\argmax}{argmax} 

\title{\LARGE \bf An updated look on the convergence and consistency of data-driven dynamical models}
\author{Kristian Løvland$^{1, 2, *}$, Bjarne Grimstad$^{1, 2}$ and Lars Struen Imsland$^{1}$\thanks{*: kristian.lovland@ntnu.no}\thanks{1: Norwegian University of Science and Technology, Trondheim, Norway.}\thanks{2: Solution Seeker AS, Oslo, Norway.}
\ifdefined\ONLINEVERSION
\thanks{© 2024 IEEE. Personal use of this material is permitted. Permission
from IEEE must be obtained for all other uses, in any current or future
media, including reprinting/republishing this material for advertising or
promotional purposes, creating new collective works, for resale or
redistribution to servers or lists, or reuse of any copyrighted
component of this work in other works.}
\fi
}

\begin{document}
\maketitle
\thispagestyle{empty}
\pagestyle{empty}

\begin{abstract}
Deep sequence models are receiving significant interest in current machine learning research. By representing probability distributions that are fit to data using maximum likelihood estimation, such models can model data on general observation spaces (both continuous and discrete-valued). Furthermore, they can be applied to a wide range of modelling problems, including modelling of dynamical systems which are subject to control. The problem of learning data-driven models of systems subject to control is well studied in the field of system identification. In particular, there exist theoretical convergence and consistency results which can be used to analyze model behaviour and guide model development. However, these results typically concern models which provide point predictions of continuous-valued variables. Motivated by this, we derive convergence and consistency results for a class of nonlinear probabilistic models defined on a general observation space. The results rely on stability and regularity assumptions, and can be used to derive consistency conditions and bias expressions for nonlinear probabilistic models of systems under control. We illustrate the results on examples from linear system identification and Markov chains on finite state spaces.
\end{abstract}

\section{Introduction}
In recent years, a significant amount of resources in the field of machine learning has been directed towards the development of sequence models. A particularly noteworthy example is the emergence of large language models, which model text \cite{zhao2023survey, min2023recent}. However, sequence models have also found applications in other domains, such as time series modelling \cite{lim2021time, wen2023transformers}. This enables their application for data-driven modelling of dynamical systems.

Some important sequence model architectures which have emerged are the long short-term memory \cite{hochreiter1997long}, gated recurrent unit \cite{cho2014learning}, temporal convolutional networks \cite{lea2016temporal}, models based on the transformer \cite{vaswani2017attention}, and deep models based on linear state-space models \cite{gu2021efficiently, smith2022simplified, gu2023mamba}. These models differ in their design, but they do have some commonalities. Two important properties commonly held by such models is that they are based on deep neural networks (which means that they can model nonlinear relationships), and that they model \textit{probability distributions} over a wide range of data modalities and observation spaces \cite{lim2021time}.

We are interested in the application of sequence models to \textit{controlled} dynamical systems, where we interpret the ``control" term quite broadly; some examples of applications which inspire our work are the use of data-driven models in control theory and system identification \cite{ljung_system_1999, pillonetto2023deep}, model-based reinforcement learning \cite{moerland2023model}, and more broadly, the use of data-driven models in decision making strategies based on \textit{planning} \cite{russell2010artificial}.

In all of these problems, the objective is to choose values of an input $u$, or a strategy for choosing $u$, which are likely to result in desired values of an output $y$. For a learned model to be of use, it is then crucial that it captures the effect on $y$ of actively intervening on $u$. Using control theoretic terms, we want to model the \textit{open-loop} behaviour of the system\cite{ljung_system_1999}; an alternative perspective is that we want to learn the \textit{causal effect} of $u$ on $y$ (see \cite{pearl_causality_2009} for an in-depth discussion of learning causal effects from data).

For linear dynamical systems, it is known that open-loop system behaviour can be hard to learn when data are gathered under feedback control, due to spurious correlations between $u$ and $y$ which are introduced by the controller. This problem is well-studied in the field of system identification \cite{ljung_system_1999}, and there exist many theoretical results which provide consistency conditions, and which quantify model bias \cite{van_den_hof_closed-loop_1998, forssell_closed-loop_1999}. While the main body of literature concerns linear systems, results also exist for nonlinear systems \cite{ljung_convergence_1978, caines_stationary_1978}. As pointed out by \cite{pillonetto2023deep}, however, the classical theory of system identification (both linear and nonlinear) strongly relies on the \textit{Prediction Error} (PE) identification framework, which assumes that models provide point predictions. To our knowledge, there has been relatively little focus in the field on models which provide predictive probability distributions. A similar observation is made in \cite{pillonetto2023deep}.

Still, we believe that the statistical challenges caused by control should be expected to arise for more general classes of controlled systems. We also believe that the questions of \textit{convergence} and \textit{consistency}, which are central in system identification, are relevant in the context of deep sequence models. Motivated by this, we derive an asymptotic convergence result for a class of nonlinear input-driven probabilistic sequence models. The result relies on fairly basic regularity and stability assumptions, and can be used to derive consistency conditions and bias terms. We demonstrate the applicability of the result by deriving a consistency condition for the class of $s$-th order Markov systems, which are characterized by their finite-length dependence on the past.

Our convergence result along with its proof is inspired by the classical result of \cite{ljung_convergence_1978}, while the problem statement and consistency result take their main inspiration from deep probabilistic sequence models. All of our results apply to systems with both continuous and discrete-valued input and output variables, and they apply to models which provide predictive distributions, in addition to models that provide point predictions. Furthermore, they apply to models whose dependence on the past may be indefinitely long, which is generally required to describe models which utilize a hidden state. Inspired by current practices in deep learning, we assume that learning is done through \textit{maximum likelihood} (ML) estimation, and we choose not to focus on the problem of parameter identification, since the model classes which motivate our work are typically over-parameterized.

Our goal is to present results which feel relevant and accessible to practitioners of both machine learning and control theory. To this end, we consider a quite simple problem formulation, and prove convergence in a manner which draws as much inspiration from the classical system identification result of \cite{ljung_convergence_1978} as it does from more general convergence and consistency results which are available to us, e.g. through \cite{van2008hidden, douc_consistency_2011}. We believe that this can contribute towards the discussion and analysis of control theoretic questions which arise in practice, but which are absent from much of the literature on statistical learning, and machine learning in particular.

\subsection{Structure of paper}
We start by summarizing related work in Section \ref{sec:related-work}, before we proceed in Section \ref{sec:problem-statement} by stating our learning problem, followed by our stability and regularity assumptions in Section \ref{sec:assumptions}. Our main convergence result is presented in Section \ref{sec:convergence}, and its use for deriving consistency conditions is discussed and demonstrated in Section \ref{sec:consistency}. Section \ref{sec:examples} contains examples which illustrate the use of our convergence and consistency results. Section \ref{sec:conclusion} contains some concluding remarks.

\subsection{Notation}
Unless otherwise is stated, capital letters denote random variables while lowercase letters denote their realizations. The letter $t \in \{1, 2, \dots \}$ denotes time, and $X_t$ denotes the random variable $X$ at time $t$. Sequences $\{x_s, \dots, x_t \}$ may be compactly written as $x_{s:t}$. Stochastic processes $\{X_1, X_2, \dots \}$ may be written compactly as $\{ X_t \}$, and are assumed to start at $t=1$ unless else is stated. For any two distributions $p(X)$ and $q(X)$, we will denote their Kullback-Leibler divergence $\kullbp{p(X)}{q(X)}$; for conditional distributions $p(Y \condbar X)$ and $q(Y \condbar X)$ we will denote the Kullback-Leibler divergence $\kullbp{p(\cdot \condbar X)}{q(\cdot \condbar X)}$ to emphasize its dependence on the random variable $X$. We will reserve the letter $p$ to denote ``true" system distributions $p(X)$, and use $q$ to denote model distributions $q(X)$. Parameterized distributions are denoted $q_{\theta}(X)$, where $\theta$ is the parameter. The expression $\argmax_{x \in \mathcal X} f(x)$ refers to the \textit{set} of elements in $\mathcal X$ which maximize $f(x)$. We will abuse notation somewhat and write $x' = \argmax_{x \in \mathcal X} f(x)$ whenever $f(x') = \max_{x \in \mathcal X}f(x)$, even if the maximizing element of $f(x)$ is not unique.

\section{Related work}
\label{sec:related-work}
In this section, we provide a brief summary of relevant literature on estimator convergence and consistency which has inspired our work.

\paragraph{Consistency of Maximum Likelihood}
Our main technical endeavour concerns ML estimator convergence, and we also present a consistency result. Hence, our work is closely related to the question of ML estimator consistency; see e.g. \cite{wald1949note} for an influential proof of an i.i.d. data case. However, our results concern serially dependent data, and since we want them to apply to state-space types of models, we consider models which in general may condition on \textit{all} past observations (i.e. models which have infinite memory). This separates our result both from \cite{wald1949note} and from the most standard results presented in probability theory textbooks, which mainly seem to consider memoryless or finite memory models learned from ergodic stationary data \cite{newey1994large, hayashi2011econometrics}. Our emphasis on convergence as a separate question from consistency also separates us from much of the literature on ML consistency.

\paragraph{Prediction Error Identification}
The learning problem we consider can be viewed through the lens of system identification. In particular, our work is inspired by the convergence result presented in the seminal paper of Lennart Ljung on convergence of parametric prediction error identification methods \cite{ljung_convergence_1978}, as well its more recent presentation in  \cite{abdalmoaty2019identification}. Its use in deriving consistency conditions and statistical biases for the problem of closed-loop linear system identification \cite{van_den_hof_closed-loop_1998, forssell_closed-loop_1999} also serves as motivation, both for our convergence and consistency results. Furthermore, our convergence proof takes significant inspiration from \cite{ljung_convergence_1978}, as does our emphasis on convergence as a separate question from consistency. However, our result differs in that it is not restricted to the class of PE identification methods. As mentioned, this has consequences for the applicability of the results, since the PE framework is concerned with continuous-valued systems and models which provide point predictions $\hat{y}_t$ of the output $y_t$. Our results, in contrast, apply to both continuous and discrete-valued variables, and they apply to models which provide both point predictions $\hat{y}_t$ and predictive distributions $q(y_t)$. Still, we note that we do not provide a strict generalization of \cite{ljung_convergence_1978}; most notably, we will see that the generality of our observation space comes at the cost of somewhat less explicit stability conditions.

\paragraph{Hidden Markov Models}
To our knowledge, the most complete theoretical analysis of ML consistency in a context comparable to ours is presented in the context of \textit{Hidden Markov Models} (HMM), in the works of \cite{douc_asymptotics_2001, douc_asymptotic_2004, van2008hidden, douc_consistency_2011, douc_asymptotic_2012}. For our purposes, the ML consistency result of \cite{douc_consistency_2011} is particularly relevant.

Similarly to us, these works consider systems with probabilistic dynamics and general observation spaces. However, all of them consider a measure theoretic problem formulation, and their proofs rely on ergodicity assumptions and results from quite recent theory on Markov chains \cite{meyn2012markov}. Additionally, their assumptions rely on the concept of Harris recurrence. Our problem formulation, in contrast, takes its inspiration from applied machine learning; our assumptions do not rely on the notion of Harris recurrence, and our proofs do not require measure theory or ergodic theory beyond a basic ergodic theorem due to \cite{karlin_first_1975}.
\section{Problem statement}
\label{sec:problem-statement}
Our learning problem can be divided into three parts: The dataset, the model class, and the objective function. We start this section by presenting these. We then describe briefly the strategy for deriving our main result.

\subsection{Dataset}
We assume that the dataset consists of observations of an input $U$ and an output $Y$, both of which are random variables. We assume that these are generated simultaneously from a dynamical system, and that the observed data takes the form of a sequence $(U_{1:T}, Y_{1:T})$ whose distribution factorizes as follows:
\begin{equation}
\label{eq:data-generating-system}
    p(U_{1:T}, Y_{1:T}) = \prod_{t=1}^T \underbrace{p(Y_t \condbar U_{1:t-1}, Y_{1:t-1})}_{\textup{System}} \underbrace{p(U_t \condbar U_{1:t-1}, Y_{1:t})}_{\textup{Controller}}
\end{equation}
The probabilities which enter this expression can represent probability density functions (for continuous variables), probability mass functions (for variables in discrete observation spaces of finite size), or a combination of the two (sometimes called ``mixed" variables). We will refer to either of these as probability \textit{distributions}. From (\ref{eq:data-generating-system}), one can see that the input trajectory $U_{1:T}$ and output trajectory $Y_{1:T}$ are tight coupled due to feedback in the controller. This is typical for control problems, and is the source of many interesting problems concerning the dataset which is used for learning.

\subsection{Model class}
The objective is to model the system dynamics $p(Y_t \condbar U_{1:t-1}, Y_{1:t-1})$. We assume that this is done by the use of parameterized probability distributions $q_{\theta}(Y_t \condbar U_{1:t-1}, Y_{1:t-1})$ which are allowed to have the same dependencies as the true system dynamics. Denoting the set of admissible parameters $\Theta$, the model class is then $\mathcal M = \{q_{\theta} : \theta \in \Theta \}$. 

Many popular machine learning methods are \textit{over-parameterized}. Somewhat simplified, this means that the model has more parameters than is necessary to fit the data. For such models, different values of the parameter $\theta$ can represent the same model.  Inspired by this, we will not assume that there exist a unique ``true" parameter $\theta^{\star} \in \Theta$ which describes a given distribution $q \in \mathcal M$. As a consequence, the values of the parameters themselves will not be of particular interest to us.

It is worth emphasizing that this problem formulation allows models in $\mathcal M$ to have dependencies \textit{arbitrarily} far into the past. This is desirable because many models of interest, including most models which utilize a hidden state (like recurrent neural networks or architectures based on linear state-space models), have this property.

Finally, we note that even though the models as stated here are probabilistic, they may be applied in a deterministic manner by adding a random noise term.

\subsection{Objective function}
We consider learning methods in which the model is learned from data through maximization of the \textit{log-likelihood} objective function, which for each timestep $t$ is defined as:
\begin{equation}
    \ell_t(\theta) = \log q_{\theta}(Y_t \condbar U_{1:t-1}, Y_{1:t-1}) \label{eq:log-likelihood}
\end{equation}
For a dataset $(U_{1:T}, Y_{1:T})$ of length $T$, the objective is the average log-likelihood:
\begin{equation}
    L_T(\theta) = \frac{1}{T} \sum_{t=1}^T \ell_t(\theta)
\end{equation}
Our focus is on models which maximize this objective, i.e. models $q_{\hat{\theta}_T} \in \mathcal M$ for which:
\begin{equation}
\label{eq:ml-estimator}
    \hat{\theta}_T = \argmax_{\theta \in \Theta}L_T(\theta)
\end{equation}
We will call such models \textit{maximum likelihood} estimators, or ML estimators for short.

The log-likelihood objective is ubiquitous in machine learning, and it can be used to derive many common identification criteria used in system identification. We do however note that it is straight-forward to extend our later results to a larger class of objective functions, and that our focus on the log-likelihood objective is mainly done for ease of exposition. We also note that in many settings of interest (like in deep learning), model estimators rarely satisfy (\ref{eq:ml-estimator}) exactly (typical reasons being the high dimensionality of the parameter space and and the non-convexity of the learning problem). In such settings, it is perhaps most helpful to interpret (\ref{eq:ml-estimator}) as a theoretical approximation to the actual model estimator, or as a hypothetical model which bounds its best-case performance.

\subsection{Goal}
\label{sec:goal}
Our main question concerns the asymptotic behaviour of the ML estimator: \textit{What happens to $q_{\hat{\theta}_T}$ when $T \rightarrow \infty$?} The method for analyzing this will consist of two steps:
\begin{enumerate}
    \item Show that $L_T(\theta)$ converges uniformly to a limit $\overline{L}(\theta)$
    \item Conclude that $\hat{\theta}_T \rightarrow \argmax_{\theta \in \Theta} \overline{L}(\theta)$
\end{enumerate}
Step 2 is a parameter convergence result. If we let $\bar{\theta} = \argmax_{\theta \in \Theta} \overline{L}(\theta)$, we can interpret it as a model convergence result: $q_{\hat{\theta}_T} \rightarrow q_{\bar{\theta}}$. If $\overline{L}(\theta)$ lends itself to analysis, this result can be used to analyze the properties of existing learning algorithms, and to aid the design of new ones.

As we will see, step 2 follows directly from step 1, provided the convergence in step 1 is uniform. The main challenge will thus be to prove step 1.

\section{Assumptions}
\label{sec:assumptions}
For i.i.d. data, the pointwise convergence of $L_T(\theta)$ to a limit $\overline{L}(\theta)$ can typically be proved using quite standard \textit{laws of large numbers} (LLN). Under regularity assumptions, pointwise convergence can be strengthened to uniform convergence, proving the desired result \cite{hayashi2011econometrics}.

When data are generated from a dynamical system, they will generally not be i.i.d. A commonly used substitute for the i.i.d. assumption in such settings is the assumption that the data generating process is \textit{ergodic stationary} \cite{hayashi2011econometrics}.

Briefly put, a stochastic process is stationary if its distribution does not change with time, and ergodic if is asymptotically independent with itself. For ergodic stationary processes, all statistical behaviour can be inferred from a single, infinitely long trajectory. As a consequence, they follow laws of large numbers. In their simplest form, such laws state that the time average $\frac{1}{T}\sum_{t=1}^T X_t$ of the stochastic process $\{ X_t \}$ converges to a limit $\bar{X}$ (in some stochastic sense) when $T \rightarrow \infty$. However, ergodic stationarity also gives rise to LLNs which guarantee the convergence of time averages of functions $f(X_t)$ as well as time averages of functions $f(X_t, \dots, X_{t+s})$ over finite subsequences of $\{ X_t \}$ \cite{karlin_first_1975}.

To ensure that the limit $\overline{L}(\theta)$ exists, we will assume that the stochastic process $\{ U_t, Y_t \}$ is stationary. Under this assumption, ergodicity is arguably a natural assumption, since our asymptotic results by their definition rely on the behaviour of a single trajectory $(U_{1:T}, Y_{1:T})$. We do however note that if we did not assume stationarity, we could still analyze convergence under stability assumptions (this is e.g. done in \cite{ljung_convergence_1978}).

If we assume that $\{ U_t, Y_t \}$ is ergodic stationary, the aforementioned LLNs concerning functions of ergodic stationary processes seem like a good option for analyzing the convergence of $L_T(\theta)$. There is however a problem: As one can see from (\ref{eq:log-likelihood}), we are allowing $\ell_t(\theta)$ to rely on \textit{all} past information. Hence, the functions whose convergence we are studying are not restricted to finite subsequences of $\{U_t, Y_t\}$. 

To be able to use LLNs for ergodic stationary sequences, we need to make sure that the dependence of $\ell_t(\theta)$ on the past is weak enough. In a sense, the model needs to ``forget" its past sufficiently quickly.  In this section, we state precisely the sense in which this forgetting is required to happen for $L_T(\theta)$ to converge. We also state regularity assumptions which ensure that this convergence is uniform.

\subsection{Stability}
It was pointed out by \cite{abdalmoaty2019identification} in the context of system identification that forgetting is a property of stable systems. This is evident in the literature on estimator consistency. In the PE identification setting of \cite{ljung_convergence_1978}, it is for instance shown that exponential stability of all models in the model class along with stationarity of the data-generating process (i.e. the closed-loop system) ensures that the loss function converges uniformly and almost surely to a limit. For the case of HMMs, \cite{douc_asymptotics_2001, van2008hidden} show that conditions which ensure filter stability also imply uniform and almost sure convergence of a log-likelihood objective function. In our case, one could then expect the convergence of $L_T(\theta)$ to depend on the stability of the system as well as the stability of all models in $\mathcal M$.

For observation spaces as general as the ones we consider, definitions of stability will necessarily be quite abstract. Arguably, this conflicts with our goal of presenting accessible results. Therefore, we proceed by describing a notion of stability which is not defined in terms of the observation spaces of $U$ or $Y$.

The key observation which underlies our stability condition comes from the convergence proofs presented in \cite{ljung_convergence_1978} and \cite{van2008hidden}. The proofs of both results rely on stability, but they also contain an intermediate step where one shows a property which in \cite{van2008hidden} was called ``finite memory approximation". This property, which states that the effect of observations on the objective function diminishes over time, is interesting from a practical viewpoint because it applies to the objective function rather than to the observations. Since the log-likelihood function is scalar and real-valued regardless of the observation space, this property applies to general observation spaces while demanding only basic mathematical prerequisites.

We will use this property directly as our model stability assumption. To this end, we define the \textit{finite memory approximation} of the log-likelihood:
\begin{equation}
    \ell_{t, s}(\theta) := \log q_{\theta}(Y_t \condbar U_{t-s:t-1}, Y_{t-s:t-1})
\end{equation}
As the name suggests, this log-likelihood expression is an approximation of $\ell_{t}(\theta)$. It applies to the same model $q_{\theta}$, but limits the data which are available to the model to the last $s$ observations. As a consequence, $\ell_{t, s}(\theta)$  is independent of $(U_{1:t-s-1}, Y_{1:t-s-1})$. Intuitively, $|\ell_{t, s}(\theta) - \ell_t(\theta)|$ should be low when $q_{\theta}(Y_t \condbar U_{t-s:t-1}, Y_{t-s:t-1})$ is a good approximation of $q_{\theta}(Y_t \condbar U_{1:t-1}, Y_{1:t-1})$ and high when it is not.

We will see that convergence of $L_T(\theta)$ can be proved if we assume the following:
\begin{equation}
\label{eq:finite-memory-approximability-square}
\tag{S1}
\expectationp{| \ell_t(\theta) - \ell_{t, s}(\theta) |^2} \leq C \lambda^s
\end{equation}
for all $0 < s < t$, $\theta \in \Theta$, where $C < \infty$, $0 < \lambda < 1$.

When (\ref{eq:finite-memory-approximability-square}) is satisfied, the effect of observations on the model prediction diminishes with time; in a sense, $q_{\theta}$ ``forgets" its past. Measured by $| \ell_t(\theta) - \ell_{t, s}(\theta) |^2$, the expected rate of forgetting is exponential, and it happens uniformly over $\Theta$ (meaning $C$ and $\lambda$ are independent of $\theta$).

The expectation in (\ref{eq:finite-memory-approximability-square}) is taken over the closed-loop data distribution $p(U_{1:T}, Y_{1:T})$, which means that the assumption (\ref{eq:finite-memory-approximability-square}) concerns both the model class and the data-generating system. An advantage of using such a joint system and model class assumption is that it is compact and easy to state. A drawback is that it can be somewhat difficult to interpret. In practice, one may want to prove it from more explicit assumptions.

Whether this is possible is problem dependent. For the class of nonlinear systems and deterministic nonlinear models considered in \cite{ljung_convergence_1978}, (\ref{eq:finite-memory-approximability-square}) follows from fairly standard stability assumptions.

\begin{prop}
\label{prop:r-mean-stability}
Assume that the closed-loop system $\{U_t, Y_t\}$ is $r$-mean exponentially stable with $r=4$ (def. 4.8 in \cite{abdalmoaty2019identification}). Furthermore, assume that all models $q_{\theta} \in \mathcal M$ is on the form $q_{\theta}(Y_t \condbar U_{1:t-1}, Y_{1:t-1}) = \normaldist(Y_t; f_{\theta}(U_{1:t-1}, Y_{1:t-1}), \sigma^2)$, where $\sigma > 0$ is fixed and $f_{\theta}$ is uniformly stable (def. 4.9 in \cite{abdalmoaty2019identification}). Then, (\ref{eq:finite-memory-approximability-square}) is satisfied.
\end{prop}
\begin{proof}
\ifdefined\ONLINEVERSION
The proof is shown in Appendix \ref{sec:deterministic-stability}.
\else
The proof is shown in the online supplement.
\fi

\end{proof}

\subsection{Regularity}
We will assume that $\ell_t(\theta)$ is Lipschitz continuous:
\begin{equation}
\label{eq:regularity}
\tag{R1}
|\ell_t(\theta) - \ell_t(\theta')| \leq K \| \theta - \theta' \|
\end{equation}
for all $t \geq 0$, $\theta, \theta' \in \Theta$. We will also assume that $\Theta$ is compact, and we will make the following \textit{dominance} assumption:
\begin{equation}
\label{eq:dominance}
\tag{R2}
    \expectationp{\sup_{\theta \in \Theta} \ell^2_t(\theta)} < \infty \quad \textup{for all } t > 0
\end{equation}
The rationale behind these assumptions is quite pragmatic: We assume that in practice, data and model parameters will not grow arbitrarily large. As a consequence of this, there will exist \textit{some} upper bound on the sensitivity of $\ell_t(\theta)$ to $\theta$, and there will exist a bounded subset $\Theta$ of the parameter space which can realistically be relevant to the learning problem. The boundedness of the expectation of $\ell_t^2(\theta)$ restricts the expected likelihood of the model predictions, both above and below. This will typically be satisfied in settings with non-perfect observations.

A helpful comment regarding regularity assumptions is made by \cite{ljung_convergence_1978}, which points out that regularity assumptions only concern models, which we design. We believe that our regularity assumptions are not particularly limiting; the stability assumption of (\ref{eq:finite-memory-approximability-square}) is, arguably, a more interesting subject for discussion.

\section{Convergence}
\label{sec:convergence}
We are now ready to state our convergence result.

\begin{thm}
\label{thm:convergence}
Suppose that $\Theta$ is compact, $\{ U_t, Y_t \}$ is ergodic stationary, and $\ell_t(\theta)$ satisfies (\ref{eq:finite-memory-approximability-square}), (\ref{eq:regularity}) and (\ref{eq:dominance}) for all $\theta \in \Theta$. Then, the limit
\begin{equation}
\overline{L}(\theta) = \lim_{T \rightarrow \infty} \expectationp{L_T(\theta)}
\end{equation}
exists for all $\theta \in \Theta$, and satisfies $\overline{L}(\theta) = \lim_{t \rightarrow \infty} \expectationp{\ell_t(\theta)}$. Furthermore, $L_T(\theta) \rightarrow \overline{L}(\theta)$ almost surely, uniformly in $\theta$.
\end{thm}

\begin{proof}
\ifdefined\ONLINEVERSION
The proof is shown in Appendix \ref{sec:convergence-proof}.
\else
The proof is shown in the Appendix.
\fi
\end{proof}

\begin{cor}
\label{cor:model-convergence}
Under the conditions of Theorem \ref{thm:convergence},
\begin{equation}
    \hat{\theta}_T \rightarrow \argmax_{\theta \in \Theta} \overline{L}(\theta) \quad \textup{a.s.}
\end{equation}
\end{cor}

\begin{proof}
This follows directly from the fact that the $\argmax$ and limit operations are interchangeable when convergence is uniform. For a proof of this fact, see e.g. Lemma 7.2 of \cite{van2008hidden}.
\end{proof}

These results reduce the stochastic limit behaviour of model estimators $q_{\hat{\theta}_T}$ to questions concerning the asymptotic objective function $\overline{L}(\theta)$. If characterization of the minimizers of $\overline{L}(\theta)$ is feasible, the result can be used to derive consistency conditions for the ML estimator. It can also be used to derive bias expressions.

However, the degree to which such analysis is possible is problem dependent, and the strength of the results which can be derived will largely depend on the strength of the assumptions one is willing to make. In the next section, we demonstrate the effect of one such assumption.

\section{Consistency}
\label{sec:consistency}
Assume that the dependence of the closed-loop system on its past has finite length, such that:
\begin{equation}
    p(U_t, Y_t \condbar U_{1:t-1}, Y_{1:t-1}) = p(U_t, Y_t \condbar U_{t-s:t-1}, Y_{t-s:t-1})
\end{equation}
This property, which we will call the \textit{$s$-th order Markov property}, states that all information regarding the current system state is available in the last $s$ observations. If the closed-loop system has this property, its dynamics can be described using a finite dimensional \textit{regressor}
\begin{equation}
    \Phi_t :=
    \begin{bmatrix}
    U_{t-s} & \cdots & U_{t-1} & Y_{t-s} & \cdots & Y_{t-1}    
    \end{bmatrix}
\end{equation}
which may still consist of continuous and/or discrete-valued variables. Under this definition, one can verify that the stochastic process $\{ \Phi_t \}$, is a Markov chain. If the stochastic process $\{U_t, Y_t \}$ is stationary, this Markov chain is stationary as well. The distribution $p(\Phi_t)$ is then the same for all t, and we can simply denote it $p(\Phi)$. Noting that the system dynamics of (\ref{eq:data-generating-system}) can be written as
\begin{equation}
    p(U_{1:T}, Y_{1:T}) = \prod_{t=1}^T p(Y_t \condbar \Phi_t) p(U_t \condbar \Phi_t, Y_t)
\end{equation}
we can state new convergence and consistency results.

\begin{lem}
\label{lem:consistency}
Assume that the conditions of Theorem \ref{thm:convergence} hold, and that the closed-loop system $p(U_t, Y_t \condbar U_{1:t-1}, Y_{1:t-1})$ has the $s$-th order Markov property. Further, assume that all the models in $\mathcal M$ have the $s$-th order Markov property. Then,
\begin{equation}
    \hat{\theta}_T \rightarrow \argmin_{\theta \in \Theta} \expectationbelowp{p(\Phi)}{\kullbp{p(Y \condbar \Phi)}{q_{\theta}(Y \condbar \Phi)}} \quad \textup{a.s.}
\end{equation}
\end{lem}

\begin{proof}
By Theorem \ref{thm:convergence}, we have $\overline{L}(\theta) = \lim_{T \rightarrow \infty}\expectationp{\ell_t(\theta)}$. Since $\{ \Phi_t \}$ is stationary, the stochastic process $\{ \Phi_t, Y_t \}$ is stationary as well, which implies that $\overline{L}(\theta) = \expectationp{\ell_t(\theta)} = \expectationp{\log q_{\theta}(Y \condbar \Phi)}$ for all $t > 0$, where the expectation is over the stationary distribution $p(\Phi, Y)$. Factorizing this distribution as $p(\Phi, Y) = p(\Phi) p(Y \condbar \Phi)$ and subtracting the term $\log p(Y \condbar \Phi)$ (which is independent of $\theta$), we get
\begin{align}
    \argmax_{\theta \in \Theta} \overline{L}(\theta) 
    & = \argmax_{\theta \in \Theta} \expectationbelowp{p(\Phi, Y)}{\log q_{\theta}(Y \condbar \Phi)} \\
    & = \argmin_{\theta \in \Theta} \expectationbelowp{p(\Phi)}{\expectationbelowp{p(Y \condbar \Phi)}{\frac{\log p(Y \condbar \Phi)}{\log q_{\theta}(Y \condbar \Phi)}}} \\
    & = \argmin_{\theta \in \Theta} \expectationbelowp{p(\Phi)}{\kullbp{p(Y \condbar \Phi)}{q_{\theta}(Y \condbar \Phi)}} \label{eq:convergence-kl-divergence}
\end{align}
By Corollary \ref{cor:model-convergence}, $\hat{\theta}_T \rightarrow \argmax_{\theta \in \Theta} \overline{L}(\theta)$ a.s., which finishes the proof.
\end{proof}

\begin{cor}
\label{cor:full-support-implies-consistency}
Let $p^{\star}$ denote the true open-loop system, and assume that $p^{\star} \in \mathcal M$. Let $\Theta^{\star} = \{ \theta \in \Theta : q_{\theta} = p^{\star} \}$ denote the set of all parameters which parameterize the true system. Assume that $p(\Phi)$ is supported on the whole domain of $\Phi$. Then $\hat{\theta}_T \rightarrow \Theta^{\star}$ almost surely.
\end{cor}

\begin{proof}
Fix an arbitrary $\Phi$, and let $\theta^{\star} \in \Theta^{\star}$ be a parametrization of the true system, such that $q_{\theta^{\star}} = p^{\star}$. Furthermore, let
\begin{equation}
    \hat{\theta} = \argmin_{\theta \in \Theta} \expectationbelowp{p(\Phi)}{\kullbp{p(Y \condbar \Phi)}{q_{\theta}(Y \condbar \Phi)}}
\end{equation}
By Lemma \ref{lem:consistency}, we then have $\hat{\theta}_T \rightarrow \hat{\theta}$ a.s. Furthermore,
\begin{align}
    0 & \leq
    \expectationbelowp{p(\Phi)}{\kullbp{p(Y \condbar \Phi)}{q_{\hat{\theta}}(Y \condbar \Phi)}} \\
    & \leq \expectationbelowp{p(\Phi)}{\kullbp{p(Y \condbar \Phi)}{q_{\theta^{\star}}(Y \condbar \Phi)}} = 0
\end{align}
By assumption we have $p(\Phi) > 0$, which implies that $\kullbp{p(\cdot \condbar \Phi)}{q_{\hat{\theta}}(\cdot \condbar \Phi)} = 0$. Thus, $q_{\hat{\theta}}(\cdot \condbar \Phi) = p(\cdot \condbar \Phi)$. Since $\Phi$ was arbitrary, this equality holds for all $\Phi$, which completes the proof.
\end{proof}

In other words, $p(\Phi) > 0$ for all $\Phi$ is a consistency condition. This should make sense: If all values of $\Phi$ are persistently being visited by the system, the ML estimator of $p(Y \condbar \Phi)$ will eventually become correct. 

\section{Examples}
\label{sec:examples}
In this section, we illustrate our convergence and consistency results with two simple examples which, respectively, concern systems with continuous and discrete-valued variables. First, we use Theorem \ref{thm:convergence} to derive a well-known convergence result from linear system identification. Then, we use Lemma \ref{lem:consistency} to derive a consistency condition for controlled Markov chains on finite state spaces. We conclude with some remarks regarding excitation and exploration.

\subsection{Linear system identification}
A commonly used data-generating system in system identification \cite{soderstrom_system_2001, ljung_system_1999} is the linear closed-loop system
\begin{align}
    y_t & = G(q) u_t + H(q) e_t \label{eq:linear-system-dynamics} \\
    u_t & = r_t - K(q) y_t \label{eq:linear-controller-dynamics}
\end{align}
Here, $q$ denotes the time shift operator, and functions $F(q)$ denote polynomials in $q^{-1}$ (see \cite{ljung_system_1999} for an introduction to this framework). If the polynomial $G(q)$ is strictly proper, $H(q)$ and $K(q)$ are proper, and $\{ e_t \}$ and $\{ r_t \}$ are stationary stochastic processes, this system gives rise to a distribution described by (\ref{eq:data-generating-system}). The variables $u_t$, $y_t$, $e_t$ and $r_t$ are all random, but we denote them with lowercase letters here since this is common in system identification literature.

It is common to mirror the true open-loop system dynamics and model $G_{\theta}(q)$ and $H_{\theta}(q)$, where $G_{\theta}(q)$ is strictly proper and $H_{\theta}(q)$ is proper. Under this model,
\begin{align}
    y_t & = G_{\theta}(q) u_t + H_{\theta}(q) e_t \label{eq:linear-model} \\
    \implies y_t & = H_{\theta}^{-1}(q) G_{\theta}(q) u_t + [1 - H_{\theta}^{-1}(q)]y_t + e_t
\end{align}
Let $\hat{y}_t = H_{\theta}^{-1}(q) G_{\theta}(q) u_t + [1 - H_{\theta}^{-1}(q)]y_t$. If we assume that $\{ e_t \}$ is white and Gaussian distributed with zero mean and fixed variance $\sigma^2$, this implies that
\begin{equation}
    q_{\theta}(y_t \condbar u_{1:t-1}, y_{1:t-1}) = \normaldist(y_t; \hat{y}_t, \sigma^2)
\end{equation}
which implies that the log-likelihood is given by
\begin{equation}
    \ell_t(\theta) = -\log(\sqrt{2 \pi} \sigma) -\frac{1}{\sigma^2} \varepsilon_t^2
\end{equation}
where $\varepsilon_t = y_t - \hat{y}_t$ is the \textit{prediction error}. From this, we can see that ML estimation under the above model assumptions is equivalent to prediction error minimization under the identification criterion $\frac{1}{T} \sum_{t=1}^T\varepsilon_t^2$.

It is common to represent $G_{\theta}(q)$ and $H_{\theta}(q)$ as fractions of finite order polynomials in $q^{-1}$ (e.g. using an ARMAX model). Usually, one can then verify that all models in $\mathcal M$ are \textit{uniformly stable} by verifying that the necessary poles are inside the unit circle \cite{ljung_system_1999}. If the closed-loop system has all its poles inside the unit circle, $\{ U_t, Y_t \}$ is \textit{$r$-mean exponentially stable with $r=4$} \cite{ljung_convergence_1978}. By Proposition \ref{prop:r-mean-stability}, these stability conditions imply (\ref{eq:finite-memory-approximability-square}). If we assume that $\{ e_t \}$ and $\{ r_t \}$ are ergodic stationary, $\{ u_t, y_t \}$ is ergodic stationary as well, as long as the closed-loop system is stable. Assuming that the model regularity conditions (\ref{eq:regularity}) and (\ref{eq:dominance}) hold, and that $\Theta$ is compact, Theorem \ref{thm:convergence} then implies that
\begin{equation}
    \hat{\theta}_T \rightarrow \argmin_{\theta \in \Theta} \expectationp{\varepsilon_t^2}
\end{equation}
which is a well-known result in system identification (see e.g. Chapter 8 of \cite{ljung_system_1999}).


\subsection{Markov chain with finite state space}
Let $S_t = Y_t$ denote a \emph{state}, let $A_t = U_t$ denote an \emph{action}, and assume that their joint trajectory can be factorized as
\begin{equation}
    p(S_{1:T}, A_{1:T}) = \prod_{t=1}^{T} \underbrace{p(S_t \condbar S_{t-1}, A_{t-1})}_{\textup{State dynamics}} \underbrace{p(A_t \condbar S_t)}_{\textup{Policy}}
\end{equation}
This stochastic process, which is commonly used to describe system dynamics in \textit{Reinforcement Learning} (RL) \cite{sutton_reinforcement_2018}, is an instance of (\ref{eq:data-generating-system}). In the context of RL, the state spaces of $S_t$ and $A_t$ are often finite, (this may e.g. be the case when $S_t$ represents the state of a board game). If we define $\Phi_t = (S_{t-1}, A_{t-1})$, the dynamics can be rewritten as
\begin{equation}
    p(S_{1:T}, A_{1:T}) = \prod_{t=1}^{T} p(S_t \condbar \Phi_t) p(A_t \condbar S_t)
\end{equation}
One can see that $\{ \Phi_t \}$ is a Markov chain. Thus, (\ref{eq:finite-memory-approximability-square}) is trivially satisfied, since $\ell_t(\theta) = \ell_{t, 1}(\theta)$. For finite state spaces, (\ref{eq:dominance}) is true if the probability mass functions provided by all models in $\mathcal M$ have a uniform lower bound. If we assume that the Markov chain $\{ \Phi_t \}$ is ergodic, $\Theta$ is compact and that the model is parameterized in such a way that (\ref{eq:regularity}) holds, the conditions of Theorem \ref{thm:convergence} are satisfied. Since $\{ \Phi_t \}$ is a Markov chain, the conditions of Lemma \ref{lem:consistency} are then satisfied as well. Lemma \ref{lem:consistency} states that the model becomes consistent as long as the stationary distribution $p(\Phi)$ satisfies $p(\Phi) > 0$ for all $\Phi > 0$. Thus, the RL convergence criterion of all state-action pairs being visited infinitely often \cite{sutton_reinforcement_2018} is also a consistency condition for the system model $q_{\theta}(Y_t \condbar \Phi_t)$.


Corollary \ref{cor:full-support-implies-consistency} can be applied to linear systems as well. For instance, representing (\ref{eq:linear-model}) by an ARX model gives rise to the model $y_t = \theta^T \Phi_t + e_t$, where $\Phi_t$ is defined as above \cite{ljung_system_1999}. If we assume that the true open-loop dynamics as well as the controller dynamics can be represented in this way, it is evident from (\ref{eq:linear-system-dynamics}) and (\ref{eq:linear-controller-dynamics}) that we can write $\Phi_{t+1} = M \Phi_t + N [e_t \; r_t]^T$, where $M$ and $N$ are suitable block matrices. If we assume that $M$ is stable and that $\{ e_t \}$ and $\{ r_t \}$ are white and Gaussian, the stationary distribution of $\{ \Phi_t \}$ will be Gaussian as well. In such a case, $p(\Phi) > 0$ for all $\Phi$ if and only if $\expectationp{\Phi \Phi^T} \succ 0$. That is, $p(\Phi) > 0$ for all $\Phi$ if and only if the signal $\{ \Phi_t \}$ is \textit{persistently exciting}.

\section{Concluding remarks}
\label{sec:conclusion}
We have derived convergence and consistency results for a class of nonlinear probabilistic models of controlled systems. Under stability and regularity conditions which draw inspiration from literature on system identification and Hidden Markov Models, the results enable theoretical analysis of sequence models which are currently emerging in the machine learning literature.

A natural direction for further research is to apply the convergence result of Theorem \ref{thm:convergence} to new problems. The question of stability in a general observation space, as illustrated by the stability condition (\ref{eq:finite-memory-approximability-square}), is also an interesting subject for further investigation.


\bibliographystyle{IEEEtran}
\bibliography{bibliography} 

\begin{thebibliography}{10}
\providecommand{\url}[1]{#1}
\csname url@samestyle\endcsname
\providecommand{\newblock}{\relax}
\providecommand{\bibinfo}[2]{#2}
\providecommand{\BIBentrySTDinterwordspacing}{\spaceskip=0pt\relax}
\providecommand{\BIBentryALTinterwordstretchfactor}{4}
\providecommand{\BIBentryALTinterwordspacing}{\spaceskip=\fontdimen2\font plus
\BIBentryALTinterwordstretchfactor\fontdimen3\font minus
  \fontdimen4\font\relax}
\providecommand{\BIBforeignlanguage}[2]{{%
\expandafter\ifx\csname l@#1\endcsname\relax
\typeout{** WARNING: IEEEtran.bst: No hyphenation pattern has been}%
\typeout{** loaded for the language `#1'. Using the pattern for}%
\typeout{** the default language instead.}%
\else
\language=\csname l@#1\endcsname
\fi
#2}}
\providecommand{\BIBdecl}{\relax}
\BIBdecl

\bibitem{zhao2023survey}
W.~X. Zhao, K.~Zhou, J.~Li, T.~Tang, X.~Wang, Y.~Hou, Y.~Min, B.~Zhang,
  J.~Zhang, Z.~Dong \emph{et~al.}, ``A survey of large language models,''
  \emph{arXiv preprint arXiv:2303.18223}, 2023.

\bibitem{min2023recent}
B.~Min, H.~Ross, E.~Sulem, A.~P.~B. Veyseh, T.~H. Nguyen, O.~Sainz, E.~Agirre,
  I.~Heintz, and D.~Roth, ``Recent advances in natural language processing via
  large pre-trained language models: A survey,'' \emph{ACM Computing Surveys},
  vol.~56, no.~2, pp. 1--40, 2023.

\bibitem{lim2021time}
B.~Lim and S.~Zohren, ``Time-series forecasting with deep learning: a survey,''
  \emph{Philosophical Transactions of the Royal Society A}, vol. 379, no. 2194,
  p. 20200209, 2021.

\bibitem{wen2023transformers}
Q.~Wen, T.~Zhou, C.~Zhang, W.~Chen, Z.~Ma, J.~Yan, and L.~Sun, ``Transformers
  in time series: a survey,'' in \emph{Proceedings of the Thirty-Second
  International Joint Conference on Artificial Intelligence}, 2023, pp.
  6778--6786.

\bibitem{hochreiter1997long}
S.~Hochreiter and J.~Schmidhuber, ``Long short-term memory,'' \emph{Neural
  computation}, vol.~9, no.~8, pp. 1735--1780, 1997.

\bibitem{cho2014learning}
K.~Cho, B.~Van~Merri{\"e}nboer, C.~Gulcehre, D.~Bahdanau, F.~Bougares,
  H.~Schwenk, and Y.~Bengio, ``Learning phrase representations using rnn
  encoder-decoder for statistical machine translation,'' \emph{arXiv preprint
  arXiv:1406.1078}, 2014.

\bibitem{lea2016temporal}
C.~Lea, R.~Vidal, A.~Reiter, and G.~D. Hager, ``Temporal convolutional
  networks: A unified approach to action segmentation,'' in \emph{Computer
  Vision--ECCV 2016 Workshops: Amsterdam, The Netherlands, October 8-10 and
  15-16, 2016, Proceedings, Part III 14}.\hskip 1em plus 0.5em minus
  0.4em\relax Springer, 2016, pp. 47--54.

\bibitem{vaswani2017attention}
A.~Vaswani, N.~Shazeer, N.~Parmar, J.~Uszkoreit, L.~Jones, A.~N. Gomez,
  {\L}.~Kaiser, and I.~Polosukhin, ``Attention is all you need,''
  \emph{Advances in neural information processing systems}, vol.~30, 2017.

\bibitem{gu2021efficiently}
A.~Gu, K.~Goel, and C.~R{\'e}, ``Efficiently modeling long sequences with
  structured state spaces,'' \emph{arXiv preprint arXiv:2111.00396}, 2021.

\bibitem{smith2022simplified}
J.~T. Smith, A.~Warrington, and S.~W. Linderman, ``Simplified state space
  layers for sequence modeling,'' \emph{arXiv preprint arXiv:2208.04933}, 2022.

\bibitem{gu2023mamba}
A.~Gu and T.~Dao, ``Mamba: Linear-time sequence modeling with selective state
  spaces,'' \emph{arXiv preprint arXiv:2312.00752}, 2023.

\bibitem{ljung_system_1999}
L.~Ljung, ``\BIBforeignlanguage{en}{System {Identification}},'' in
  \emph{\BIBforeignlanguage{en}{Wiley {Encyclopedia} of {Electrical} and
  {Electronics} {Engineering}}}.\hskip 1em plus 0.5em minus 0.4em\relax John
  Wiley \& Sons, Ltd, 1999, pp. 1--19.

\bibitem{pillonetto2023deep}
G.~Pillonetto, A.~Aravkin, D.~Gedon, L.~Ljung, A.~H. Ribeiro, and T.~B.
  Sch{\"o}n, ``Deep networks for system identification: a survey,'' \emph{arXiv
  preprint arXiv:2301.12832}, 2023.

\bibitem{moerland2023model}
T.~M. Moerland, J.~Broekens, A.~Plaat, C.~M. Jonker \emph{et~al.},
  ``Model-based reinforcement learning: A survey,'' \emph{Foundations and
  Trends{\textregistered} in Machine Learning}, vol.~16, no.~1, pp. 1--118,
  2023.

\bibitem{russell2010artificial}
S.~J. Russell and P.~Norvig, \emph{Artificial intelligence a modern
  approach}.\hskip 1em plus 0.5em minus 0.4em\relax London, 2010.

\bibitem{pearl_causality_2009}
J.~Pearl, \emph{Causality}.\hskip 1em plus 0.5em minus 0.4em\relax Cambridge
  University Press, Sep. 2009.

\bibitem{van_den_hof_closed-loop_1998}
P.~Van~den Hof, ``Closed-loop issues in system identification,'' \emph{Annual
  Reviews in Control}, vol.~22, pp. 173--186, Jan. 1998.

\bibitem{forssell_closed-loop_1999}
U.~Forssell and L.~Ljung, ``\BIBforeignlanguage{en}{Closed-loop identification
  revisited},'' \emph{\BIBforeignlanguage{en}{Automatica}}, vol.~35, no.~7, pp.
  1215--1241, Jul. 1999.

\bibitem{ljung_convergence_1978}
L.~Ljung, ``Convergence analysis of parametric identification methods,''
  \emph{IEEE Transactions on Automatic Control}, vol.~23, no.~5, pp. 770--783,
  Oct. 1978.

\bibitem{caines_stationary_1978}
P.~Caines, ``Stationary linear and nonlinear system identification and
  predictor set completeness,'' \emph{IEEE Transactions on Automatic Control},
  vol.~23, no.~4, pp. 583--594, Aug. 1978.

\bibitem{van2008hidden}
\BIBentryALTinterwordspacing
R.~van Handel, ``Hidden markov models,'' \emph{Lecture notes}, 2008. [Online].
  Available: \url{https://web.math.princeton.edu/~rvan/orf557/hmm080728.pdf}
\BIBentrySTDinterwordspacing

\bibitem{douc_consistency_2011}
R.~Douc, E.~Moulines, J.~Olsson, and R.~v. Handel, ``Consistency of the maximum
  likelihood estimator for general hidden {Markov} models,'' \emph{The Annals
  of Statistics}, vol.~39, no.~1, pp. 474--513, Feb. 2011.

\bibitem{wald1949note}
A.~Wald, ``Note on the consistency of the maximum likelihood estimate,''
  \emph{The Annals of Mathematical Statistics}, vol.~20, no.~4, pp. 595--601,
  1949.

\bibitem{newey1994large}
W.~K. Newey and D.~McFadden, ``Large sample estimation and hypothesis
  testing,'' \emph{Handbook of econometrics}, vol.~4, pp. 2111--2245, 1994.

\bibitem{hayashi2011econometrics}
F.~Hayashi, \emph{Econometrics}.\hskip 1em plus 0.5em minus 0.4em\relax
  Princeton University Press, 2011.

\bibitem{abdalmoaty2019identification}
M.~Abdalmoaty, ``Identification of stochastic nonlinear dynamical models using
  estimating functions,'' Ph.D. dissertation, KTH Royal Institute of
  Technology, 2019.

\bibitem{douc_asymptotics_2001}
R.~Douc and C.~Matias, ``Asymptotics of the {Maximum} {Likelihood} {Estimator}
  for {General} {Hidden} {Markov} {Models},'' \emph{Bernoulli}, vol.~7, no.~3,
  pp. 381--420, 2001.

\bibitem{douc_asymptotic_2004}
R.~Douc, E.~Moulines, and T.~Ryd\'en, ``Asymptotic properties of the maximum
  likelihood estimator in autoregressive models with {Markov} regime,''
  \emph{The Annals of Statistics}, vol.~32, no.~5, pp. 2254--2304, Oct. 2004.

\bibitem{douc_asymptotic_2012}
R.~Douc and E.~Moulines, ``Asymptotic properties of the maximum likelihood
  estimation in misspecified hidden {Markov} models,'' \emph{The Annals of
  Statistics}, vol.~40, no.~5, pp. 2697--2732, Oct. 2012.

\bibitem{meyn2012markov}
S.~P. Meyn and R.~L. Tweedie, \emph{Markov chains and stochastic
  stability}.\hskip 1em plus 0.5em minus 0.4em\relax Springer Science \&
  Business Media, 2012.

\bibitem{karlin_first_1975}
S.~Karlin and H.~M. Taylor, \emph{\BIBforeignlanguage{en}{A first course in
  stochastic processes}}, 2nd~ed.\hskip 1em plus 0.5em minus 0.4em\relax New
  York: Academic Press, 1975.

\bibitem{soderstrom_system_2001}
T.~Söderström and P.~Stoica, \emph{System {Identification}}.\hskip 1em plus
  0.5em minus 0.4em\relax Prentice Hall International, 2001.

\bibitem{sutton_reinforcement_2018}
R.~S. Sutton and A.~G. Barto, \emph{\BIBforeignlanguage{en}{Reinforcement
  learning: an introduction}}, second edition~ed., ser. Adaptive computation
  and machine learning series.\hskip 1em plus 0.5em minus 0.4em\relax
  Cambridge, Massachusetts: The MIT Press, 2018.

\bibitem{lyons_strong_1988}
R.~Lyons, ``Strong laws of large numbers for weakly correlated random
  variables.'' \emph{Michigan Mathematical Journal}, vol.~35, no.~3, pp.
  353--359, Jan. 1988.

\end{thebibliography}

\ifdefined\ONLINEVERSION
\appendices
\section{Proof of Theorem \ref{thm:convergence}}
\label{sec:convergence-proof}
We first state two results which will be useful.
\begin{prop}
\label{prop:ergodic-stationary-function}
Let $\{ X_t \}$ be an ergodic stationary process, and let $s < \infty$ be a constant. Then, the process $\{ f(X_{t-s}, \dots, X_t) \}$, which starts at $t = s+1$, is ergodic stationary as well.
\end{prop}

\begin{proof}
This follows from Theorem 5.6 in \cite{karlin_first_1975}, which states that if $\{ X_t \}$ is an ergodic stationary (e.s.) process, the process $\{ f(X_{t}, \dots, X_{t+s}) \}$ is e.s. as well. Since the process $\{ X_{t} \}$ (which starts at $t=1$) is e.s., the process $\{ X_{t-s} \}$ (starting at time $t = s+1$) is also e.s. Thus, the process $\{ f(X_{t-s}, \dots, X_t) \}$ (starting at time $t=s+1$) is ergodic stationary.
\end{proof}

\begin{prop}
\label{prop:summable-autocovariance-lln}
Let $\{ X_t \}$ be a zero-mean stochastic process which satisfies
\begin{equation}
    | \expectationp{X_{t-s} X_t } | \leq C \lambda^s
\end{equation}
for all $t > 0, s \geq 0$. Then, $\frac{1}{T}\sum_{t=1}^T X_t \rightarrow 0$ almost surely.
\end{prop}

\begin{proof}
This result appears in different forms throughout the statistics literature. If we define $\Phi_1(n) = C \lambda^n$ and normalize $X_t$ by $\sqrt{C}$, the desired result follows directly from Corollary 11 of \cite{lyons_strong_1988}.
\end{proof}

We proceed by proving Theorem \ref{thm:convergence} in three steps. We emphasize that Step 1 and 3 closely follow the proof of Proposition 7.5 in \cite{van2008hidden}, while Step 2 is similar to the first part of the proof of Lemma 3.1 in \cite{ljung_convergence_1978}.

\paragraph{Step 1 -- Existence of $\overline{L}(\theta)$}
We note first that by Jensen's inequality, (\ref{eq:finite-memory-approximability-square}) implies that
\begin{equation}
\label{eq:finite-memory-approximability}
\tag{S2}
\expectationp{| \ell_t(\theta) - \ell_{t, s}(\theta) |} \leq C \lambda^s
\end{equation}
for all $t > s > 0, \, \theta \in \Theta$, for constants $C < \infty$, $0 < \lambda < 1$, different from the constants of (\ref{eq:finite-memory-approximability-square}). Furthermore, we note that the process $\{ \ell_{t, s}(\theta) \}$, which starts at $t=s+1$, is ergodic stationary by Proposition \ref{prop:ergodic-stationary-function}. The stationarity of $\{ \ell_{t, s}(\theta) \}$ implies that $\expectationp{\ell_{t, s}(\theta)}$ is the same for all $t > s$; in particular, we have $\expectationp{\ell_{s, s-1}(\theta)} = \expectationp{\ell_{t, s}(\theta)}$ for all $t > s$. From the definition of $\ell_t(\theta)$ and $\ell_{t, s}(\theta)$, we can also see that $\ell_s(\theta) = \ell_{s, s-1}(\theta)$ for all $s > 0$.

Let $\mu_t := \expectationp{\ell_t(\theta)}$ be the expected log-likelihood at time $t$. For any $0 < s < t$, we then have $\mu_s = \expectationp{\ell_s(\theta)} = \expectationp{\ell_{s, s-1}(\theta)} =  \expectationp{\ell_{t, s}(\theta)}$. In particular, we have $\mu_t = \expectationp{\ell_{t+n, t}(\theta)}$ for $t, n > 0$, and by (\ref{eq:finite-memory-approximability}) we have
\begin{align}
    |\mu_{t+n} - \mu_t| 
    & = | \expectationp{\ell_{t+n}(\theta) - \ell_{t+n, t}(\theta)} | \\
    & \leq \expectationp{| \ell_{t+n}(\theta) - \ell_{t+n, t}(\theta) |} \\
    & \leq C \lambda^t
\end{align}
Thus, $\sup_{n \geq 0} |\mu_{t+n} - \mu_t| \rightarrow 0$ as $t \rightarrow \infty$, which means that $\{ \mu_t \}$ is a Cauchy sequence, and hence, convergent. We denote its limit $\overline{\ell}(\theta)$. Since it consist of Cesàro sums of $\mu_t$, the sequence $\{\expectationp{L_T(\theta)}\} = \{\frac{1}{T}\sum_{t=1}^T \mu_t\}$ then converges as well. Although the limit of this sequence is also $\overline{\ell}(\theta)$, we will denote it $\overline{L}(\theta)$.

\paragraph{Step 2 -- Convergence of $L_T(\theta)$}
We have shown that $\overline{L}(\theta)$ exists. We now proceed by showing that $L_T(\theta)$ converges to $\overline{L}(\theta)$ with probability one. For this purpose, we define
\begin{align}
    \eta_t & := \ell_t(\theta) - \expectationp{\ell_t(\theta)} \\
    \eta_{t, s} & := \ell_{t, s}(\theta) - \expectationp{\ell_t(\theta)}
\end{align}
Note that we have omitted the dependence of $\eta_t$ and $\eta_{t, s}$ on $\theta$ for notational convenience, and that the following should be read as holding for any $\theta \in \Theta$.

Recall that $\ell_{t, s}(\theta)$ and $(U_{1:t-s-1}, Y_{1:t-s-1})$ are independent by the definition of $\ell_{t, s}(\theta)$. Thus, $\ell_{t, s}(\theta)$ and $\ell_{t-s-1}(\theta)$ are independent, which implies that $\eta_{t, s}$ and $\eta_{t-s-1}$ are independent. Furthermore, $\expectationp{\eta_t} = 0$ for all $t > 0$. Thus, $\expectationp{\eta_{t-s-1} \eta_{t,s}} = 0$. Exploiting this along with Cauchy-Schwarz' inequality, we get
\begin{align}
    | \expectationp{ \eta_{t-s-1} \eta_t } | & = |\expectationp{\eta_{t-s-1} \eta_t} - \expectationp{\eta_{t-s-1} \eta_{t, s}}| \\
    & = | \expectationp{\eta_{t-s-1}(\eta_t - \eta_{t, s})} | \\
    & \leq \expectationp{\eta_{t-s-1}^2}^{1/2} \expectationp{(\eta_t - \eta_{t, s})^2}^{1/2}
\end{align}
By (\ref{eq:finite-memory-approximability-square}), we can bound the last term by
\begin{align}
    \expectationp{(\eta_t - \eta_{t, s})^2} & = \expectationp{(\ell_t(\theta) - \ell_{t, s}(\theta))^2} \leq C \lambda^s
\end{align}
which means that
\begin{align}
    | \expectationp{ \eta_{t-s-1} \eta_t } | \leq  \expectationp{\eta_{t-s-1}^2}^{1/2} C \lambda^s
\end{align}
By (\ref{eq:dominance}), $\expectationp{\ell_t^2(\theta)} < \infty$ for all $\theta \in \Theta, t > 0$, which implies that $\expectationp{\eta_{t-s-1}^2} < \infty$ for all $t > s > 0$. Proposition \ref{prop:summable-autocovariance-lln} then implies that $\frac{1}{T}\sum_{t=1}^T \eta_t \rightarrow 0$ almost surely. Thus,
\begin{align}
    \frac{1}{T}\sum_{t=1}^T (\ell_t(\theta) - \expectationp{\ell_t(\theta)}) & \rightarrow 0 \quad \textup{a.s.}
\end{align}
for all $\theta$. Since $\lim_{T \rightarrow \infty} \frac{1}{T}\sum_{t=1}^T \expectationp{\ell_t(\theta)} = \overline{L}(\theta)$ exists, this implies that for all $\theta \in \Theta$,
\begin{equation}
    L_T(\theta) = \frac{1}{T} \sum_{t=1}^T \ell_{t}(\theta) \rightarrow
    \overline{L}(\theta) \quad \textup{a.s.}
\end{equation}

\paragraph{Step 3 -- Uniformity} As our last step, we show that this convergence is uniform in $\theta$.
We start with noting that by (\ref{eq:regularity}),
\begin{align}
    |L_T(\theta) - L_T(\theta')| 
    & = \left| \frac{1}{T}\sum_{t=1}^T (\ell_t(\theta) - \ell_t(\theta')) \right| \\
    & \leq \frac{1}{T} \sum_{t=1}^T |\ell_t(\theta) - \ell_t(\theta')| \\
    & \leq K \| \theta - \theta' \|
\end{align}
for all $\theta, \theta' \in \Theta$. In other words, $L_T(\theta)$ is Lipschitz continuous with Lipschitz constant $K$.

Since $\Theta$ is compact, it can be covered by finitely many balls of radius $\delta$ for any $\delta > 0$.
Denote the finite set of center points of these balls as $\Theta_{\delta} \subset \Theta$. Since every $\theta \in \Theta$ is at most distance $\delta$ away from some point in $\Theta_{\delta}$, meaning $\| \theta - \theta' \| \leq \delta$ for some $\theta' \in \Theta_{\delta}$, we have for each $T$
\begin{align}
    \sup_{\theta \in \Theta} |L_T(\theta) - \overline{L}(\theta)| 
    & \leq \delta K + \max_{\theta \in \Theta_{\delta}} |L_{T}(\theta) - \overline{L}(\theta)|
\end{align}
Since $L_T(\theta) \rightarrow \overline{L}(\theta)$ a.s. and $\Theta_{\delta}$ is a finite set,
\begin{equation}
    \limsup_{T \rightarrow \infty} \sup_{\theta \in \Theta} |L_T(\theta) - \overline{L}(\theta)| \leq \delta K \quad \textup{a.s.}
\end{equation}
Since $\delta K$ can be made arbitrarily small, this implies that
\begin{equation}
    \sup_{\theta \in \Theta} |L_T(\theta) - \overline{L}(\theta)| \rightarrow 0 \quad \textup{a.s.}
\end{equation}
In other words, $L_T(\theta) \rightarrow \overline{L}(\theta)$ uniformly in $\theta$.
\section{Proof of Proposition \ref{prop:r-mean-stability}}
\label{sec:deterministic-stability}
We will rely on repeated use of Jensen's inequality applied to sums of absolute values, which states that for $p \geq 1$
\begin{equation}
    \left(\sum_{i=1}^n |a_i| \right)^p \leq \sum_{i=1}^n |a_i|^p 
\end{equation}
We will also use Cauchy-Schwarz' inequality for expectations, which states that
\begin{equation}
    |\expectationp{X Y}| \leq \expectationp{X^2}^{1/2} \expectationp{Y^2}^{1/2}
\end{equation}
Finally, we will rely on the notion of an \textit{$r$-mean exponentially stable} process \cite{ljung_system_1999, abdalmoaty2019identification}: For an $r$-mean exponentially stable stochastic process $\{ Y_t \}$, there exists a random variable $Y_{t, s}$ which is independent of $Y_{1:t-s}$, and where
\begin{equation}
    \expectationp{| Y_t - Y_{t, s} |^r} < C \lambda^s
\end{equation}
for $C < \infty$, $0 < \lambda < 1$.

We will assume that $U$ and $Y$ are continuous and scalar-valued; extension to the vector-valued case is straight-forward.

Define $\hat{Y}_{t, s} = f_{\theta}(U_{t-s+1:t-1}, Y_{t-s+1:t-1})$, and assume without loss of generality that $\sigma = 1 / \sqrt{2}$. Then,
\begin{align}
    |\ell_t& (\theta) - \ell_{t, s}(\theta)| \\
    & = | (Y_t - \hat{Y}_t)^2 - (Y_t - \hat{Y}_{t, s})^2 | \nonumber \\
    & = | [(Y_t - \hat{Y}_t) - (Y_t - \hat{Y}_{t, s})] \cdot [(Y_t - \hat{Y}_t) + (Y_t - \hat{Y}_{t, s})] | \nonumber \\
    & = |[-(\hat{Y}_t - \hat{Y}_{t, s})] \cdot [(Y_t - \hat{Y}_t) + (Y_t - \hat{Y}_{t, s})]| \nonumber \\
    & \leq |\hat{Y}_t - \hat{Y}_{t,s} | \cdot | (Y_t - \hat{Y}_t) + (Y_t - \hat{Y}_{t, s}) | \nonumber \\
    & \leq |\hat{Y}_t - \hat{Y}_{t,s} | \cdot | 2Y_t - \hat{Y}_t - \hat{Y}_{t, s} | 
\end{align}
Using this inequality followed by Cauchy-Schwarz' inequality for expectations and then for sums, we get
\begin{align}
    \expectation[&(\ell_t(\theta) - \ell_{t, s}(\theta))^2] \\
    & \leq \expectationp{|\hat{Y}_t - \hat{Y}_{t,s} |^2 \cdot (2 Y_t - \hat{Y}_t - \hat{Y}_{t, s})^2} \\
    & \leq \expectationp{|\hat{Y}_t - \hat{Y}_{t, s}|^4}^{1/2} \expectationp{| 2Y_t - \hat{Y_t} - \hat{Y}_{t, s}|^4}^{1/2} \\
    & \leq \expectationp{|\hat{Y}_t - \hat{Y}_{t, s}|^4}^{1/2} \left(4 \expectation[Y_t^4] + \expectation[\hat{Y}_t^4] + \expectation[\hat{Y}_{t, s}^4] \right)^{1/2}
\end{align}
Thus, (\ref{eq:finite-memory-approximability-square}) is satisfied if the following holds:
\begin{align}
    \expectationp{|\hat{Y}_t - \hat{Y}_{t, s}|^4}^{1/2} & \leq C \lambda^s \label{eq:exponential-decay-condition} \\
    \expectation[Y_t^4], \expectation[\hat{Y}_t^4], \expectation[\hat{Y}_{t, s}^4] & < \infty \label{eq:finite-fourth-moment-condition}
\end{align}
We start by noting that by uniform model stability, we have
\begin{align}
    |& \hat{Y}_t - \hat{Y}_{t,s}|  \nonumber \\
    & = |f_{\theta}(U_{1:t-1}, Y_{1:t-1}) - f_{\theta}(U_{t-s+1:t-1}, Y_{t-s+1:t-1})| \nonumber \\
    & \leq C \sum_{k=1}^{t-s} \lambda^{t-k} (|U_k| + |Y_k|)
\end{align}
By Jensen's inequality we then get
\allowdisplaybreaks
\begin{align}
    \expectationp{|\hat{Y}_t - \hat{Y}_{t, s}|^4} & \leq \expectationp{\left(C \sum_{k=1}^{t-s} \lambda^{t-k} (|U_k| + |Y_k|)\right)^4}  \nonumber \\
    & \leq \expectationp{C^4\sum_{k=1}^{t-s} \tilde{\lambda}^{(t-k)} (|U_k|^4 + |Y_k|^4)}  \nonumber \\
    & = C^4 \sum_{k=1}^{t-s} \tilde{\lambda}^{t-k} \left(\expectationp{|U_k|^4} + \expectationp{|Y_k|^4} \right)  \nonumber \\
    & = \bar{C}\sum_{k=1}^{t-s}\tilde{\lambda}^{t-k}  \nonumber \\
    & = \bar{C} \tilde{\lambda}^s \sum_{i=0}^{t-s-1} \tilde{\lambda}^i  \nonumber \\
    & \leq \frac{\bar{C}}{1 - \tilde{\lambda}} \tilde{\lambda}^s
\end{align}
where $\tilde{\lambda} = \lambda^4$ and $\bar{C} = C^4 (\expectationp{|U_k|^4} + \expectationp{|Y_k|^4})$. Redefining the constants, we get (\ref{eq:exponential-decay-condition}).

Following a similar procedure, we can see that
\begin{align}
    |\hat{Y}_t| & = |f_{\theta}(U_{1:t-1}, Y_{1:t-1})|  \nonumber \\
    & = |f_{\theta}(U_{1:t-1}, Y_{1:t-1}) - f_{\theta}(0_{1:t-1}, 0_{1:t-1}) + f_{\theta}(0_{1:t-1}, 0_{1:t-1})|  \nonumber \\
    & \leq |f_{\theta}(U_{1:t-1}, Y_{1:t-1}) - f_{\theta}(0_{1:t-1}, 0_{1:t-1})| + |f_{\theta}(0_{1:t-1}, 0_{1:t-1})|  \nonumber \\
    & \leq C \lambda^{t-k} \sum_{k=0}^{t-1} (|U_t| + |Y_t|) + C'
\end{align}
which implies that
\begin{align}
    \expectationp{|\hat{Y}_t|^4} & \leq \expectationp{\left(  C \sum_{k=0}^{t-1} \lambda^{t-k} (|U_t| + |Y_t|) + C' \right)^4}  \nonumber \\
    & \leq \expectationp{\left(  C^4 \sum_{k=0}^{t-1} \lambda^{t-k} (|U_t|^4 + |Y_t|^4) + C'^4 \right)}  \nonumber \\
    & = C^4 \sum_{k=0}^{t-1} \lambda^{t-k} (\expectationp{|U_t|^4} + \expectationp{|Y_t|^4} + C'^4)  \nonumber \\
    & = \underbrace{C (\expectationp{|U_t|^4} + \expectationp{|Y_t|^4})}_{< \infty} \underbrace{\sum_{k=0}^{t-1}\lambda^{t-k}}_{< \infty} + \underbrace{C'^4}_{< \infty}
\end{align}
By a completely analogous derivation, $\expectationp{|\hat{Y}_{t, s}|^4} < \infty$. The fact that $\expectationp{|Y_t|^4} < \infty$ follows directly from the $r$-mean exponential stability of $Y_t$. Thus, we can conclude that (\ref{eq:finite-fourth-moment-condition}) holds as well. Since the constants in (\ref{eq:finite-fourth-moment-condition}) are uniform in $\theta$, this finishes the proof.

\else
\appendix

\fi

\end{document}